\documentclass[PRA,twocolumn,preprintnumbers,superscriptaddress,amsmath]{revtex4}
\usepackage{amssymb}
\usepackage{txfonts}
\usepackage{amssymb}
\usepackage{graphicx}
\usepackage{dcolumn}
\usepackage{bm}

\begin{document}

\title{High visibility on-chip quantum interference of single surface plasmons}

\author{Yong-Jing Cai, Ming Li, Xi-Feng Ren\footnote[1]{renxf@ustc.edu.cn}, Chang-Ling Zou, Xiao Xiong, Hua-Lin Lei, Bi-Heng Liu\footnote[2]{bhliu@ustc.edu.cn}, Guo-Ping Guo\footnote[3]{gpguo@ustc.edu.cn}, and Guang-Can Guo}
\address{Key Lab of Quantum Information,
University of Science and Technology of China, CAS, Hefei, Anhui, 230026, China.}
\address{Synergetic Innovation Center of Quantum Information
$\&$ Quantum Physics, University of Science and Technology of China, Hefei, Anhui 230026, China}
\begin{abstract}
Quantum photonic integrated circuits (QPICs) based on dielectric waveguides
have been widely used in linear optical quantum computation. Recently,
surface plasmons have been introduced to this application because
they can confine and manipulate light beyond the diffraction limit.
In this study, the on-chip quantum interference of two single surface
plasmons was achieved using dielectric-loaded surface-plasmon-polariton
waveguides. The high visibility (greater than 90\%) proves the bosonic
nature of single plasmons and emphasizes the feasibility of achieving
basic quantum logic gates for linear optical quantum computation.
The effect of intrinsic losses in plasmonic waveguides with regard
to quantum information processing is also discussed. Although the
influence of this effect was negligible in the current experiment,
our studies reveal that such losses can dramatically reduce quantum
interference visibility in certain cases; thus, quantum coherence
must be carefully considered when designing QPIC devices.
\end{abstract}
\maketitle
Photonic integrated circuits (PICs), in which multiple photonic functional
components comprise a single chip, have attracted considerable attention
owing to their small footprints, scalability, reduced power consumption,
and enhanced processing stability. In addition to their wide application
in classical information processing, integrated photonic quantum logic
gates and Shor's quantum factoring algorithm have been demonstrated
on these chips \cite{Obrien08,Obrien09}; thus, they show great feasibility
and high operation fidelity. More recently, much effort has been dedicated
to surface plasmon polaritons (SPPs), which are electron density waves
excited at the interface between a metal and a dielectric material
\cite{Ebbesen03}to further condense PICs beyond the diffraction limit.
Not only can SPPs confine light at the nanoscale \cite{Sch}, they
are also useful for integrated polarization-controlling devices \cite{PBS,Polarizer}.
Studies using periodic metallic hole arrays provided the first experimental
evidence that quantum entanglement can be preserved in the photon-SPP-photon
conversion process\cite{Alt,energy,ren06}. Furthermore, the non-classical
statistics of SPPs have been demonstrated using basic quantum Hong-Ou-Mandel
(HOM) interference \cite{Hong}, in both long-range plasmonic waveguides
(weakly confining waveguide) \cite{Fuji} and sub-wavelength metal
plasmonic waveguides \cite{Rei}. These studies indicate that assembling
quantum PICs (QPICs) using plasmonic components is possible.

However, two obstacles remain that hinder the development of SPP-based
QPICs. The first is that the experimental raw visibility of the quantum
interference realized in plasmonic waveguides is below 50\% \cite{Rei},
which is the boundary between classical and quantum interference.
This low visibility is not compelling evidence that single plasmons
are usable for quantum information processing. Interference visibility
is so important that higher quantum interference visibility implies
higher operation fidelity and a higher probability of success. For
example, the HOM interference with 95\% visibility that can be achieved
in QPICs based on dielectric waveguides is used to realize quantum
controlled-NOT gates \cite{Obrien08}. Second, loss is unavoidable
in QPICs that are based on plasmonic waveguides \cite{Sch}, owing
to the absorption of metals. Although the properties of lossy quantum
channels have been recently observed and studied in the context of
free-space quantum optics \cite{Bar,Liu}, the influence of such losses
on quantum processing using QPICs remains unknown \cite{Ste,Gupta}.

The current study experimentally achieved the on-chip quantum interference
of single plasmons using dielectric-loaded plasmonic waveguides at
telecom wavelengths. The visibility was as high as $95.7\pm8.9\%$,
which unambiguously demonstrates the bosonic nature of single plasmons
and paves the way for the performance of basic quantum operations
in plasmon-based QPICs. Furthermore, an SPP waveguide might provide
a perfect testing ground for studying lossy photonic devices owing
to the relatively high loss of such a devices compared with dielectric
devices. Our analysis indicates that sub-wavelength plasmonic components
can be used as quantum devices for QPICs only when the loss effect
is carefully addressed because loss can significantly reduce quantum
interference visibility. Because loss is inevitable in waveguides,
it is necessary to consider its influence on quantum coherence when
designing QPIC structures.

\begin{figure*}[htb]
\includegraphics[width=14cm]{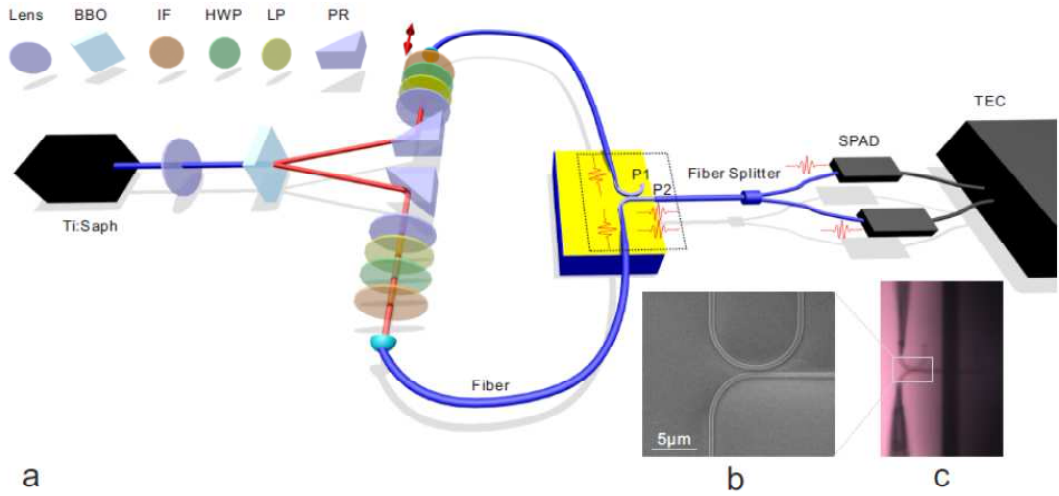} \caption{(a) Schematic of the experimental setup. Photon pairs are generated
via a degenerate type-II non-collinear spontaneous parametric down-conversion
(SPDC) process. A 1.5-W pump laser (775 nm, Coherent Inc.) is focused
on a 1-mm-long BBO crystal. The produced twin photons (1550 nm) are
separated in free space by a $6^{\circ}$ angle based on the phase-matching
condition and directed to different optical single-mode fibers. A
motorized delay line in one of the arms allows the optical-path-length
difference between the two photons to be controlled with 100-nm resolution.
Using the fiber taper connected to the single-mode fiber, the single
photons are converted into single plasmons in the plasmonic waveguide,
which then interfere with each other. We collected the photons from
P2 of the on-chip directional coupler (DC) and sent them to a second
50/50 fiber BS. Coincidence measurements revealed the quantum properties
of single plasmons. (b) A scanning electron microscope (SEM) image
of part of a typical plasmonic DC structure. (c) A CCD image showing
the coupling of the fiber taper and the SPP waveguide. A single-mode
fiber was used to collect the photons from P2 using the end-fire coupling
method.}
\end{figure*}

\section*{Results}

\textbf{The principle of quantum interference.} HOM interference,
a basic type of quantum interference that reflects the bosonic properties
of a single particle, is generally used to test the quantum properties
of single plasmons \cite{Hong}. In addition to its fundamental importance
within quantum physics, the HOM effect underlies the basic entanglement
mechanism in linear optical quantum computing \cite{Klm} because
two-qubit quantum gates, which form the core of linear optical quantum
computing, can be obtained via classical and quantum interference
(HOM interference) effects followed by a measurement-induced state
projection.

HOM interference can be described as follows: when two indistinguishable
photons enter a 50/50 beam splitter (BS) from different sides at the
same time, according to the exchange symmetry of photons (bosons),
a 50\% chance exists of obtaining two photons in output port 1 (P1);
furthermore, a 50\% chance probability exists of obtaining two photons
in output port 2 (P2). However, the two photons will never be in different
output ports. The twin photon state $|1,1\rangle$ is converted into
a quantum superposition state $1/\sqrt{2}(|2,0\rangle+|0,2\rangle)$.
This phenomenon is a signal of photon bunching and can only be explained
using a quantum mechanism \cite{Lim}. Experiments typically control
the arrival times of two photons by adjusting the path-length difference
between them and measure the photon coincidence of P1 and P2. When
two indistinguishable photons completely overlap at the BS, they give
rise to the maximum interference effect, and no coincidence exists.
Visibility is defined as $V_{1}=(C_{max}-C_{min})/C_{max}$, where
$C_{max}$ is the maximum coincidence and $C_{min}$ is the minimum
coincidence. For perfect quantum interference, $C_{min}=0$ and $V_{1}=1$,
whereas for a classical coherent laser, $V_{1}=50\%$. Consequently,
to prove that destructive interference is due to two-photon quantum
interference, the visibility must be greater than 50\%. Here, we used
a modified HOM interferometer (see Figure 1a). We collected the photons
from P2 of the first BS, sent them to the second 50/50 BS, and then
measured the coincidence. According to quantum interference theory,
a 25\% chance should exist for us to record a click when HOM interference
occurs and a 12.5\% chance otherwise. In this case, visibility was
modified as follows: $V_{2}=(C_{max}-C_{min})/C_{min}$. For perfect
quantum interference,, $C_{max}=2C_{min}$ and $V_{2}=1$. Our modified
interferometer is capable of reflecting the indistinguishability of
the input particles and can tell us whether these plasmons are bosons.

\textbf{Experimental design. }In the current experiment, we chose
a dielectric-loaded SPP waveguide (DLSPPW) \cite{Kumar} to test the
bosonic properties of the single plasmons. A DLSPPW is a typical sub-wavelength
plasmonic waveguide that is formed by placing a dielectric ridge on
top of a thin metal layer. Among the various plasmonic-waveguide structures,
DLSPPWs are promising for enriching the functional portfolio of plasmonics
owing to their dielectric-loading properties, which have been demonstrated
in practice. They can confine the lateral size of propagating modes
to the sub-wavelength scale and simultaneously transmit photons and
electrons in the same component. In addition, because the energy is
mostly confined to the surface of the metal, highly efficient control
of the waveguide-mode characteristics is possible. For example, power-monitoring
\cite{Kumar2} and switching \cite{Kala} elements with high response
speeds have been experimentally demonstrated in DLSPPWs. We used nanofabrication
techniques to prepare our plasmonic waveguide. Specifically, our waveguide
was constructed of polymethyl methacrylate (PMMA) and placed on top
of a 45-nm-thick gold layer deposited on a $SiO_{2}$ substrate. Figure
1b shows a scanning electron microscope (SEM) image of part of the
fabricated sample.

Based on our calculations, the lateral size of the single-mode DLSPPW
for photons at 1,550 nm was 600nm$\times$600nm \cite{Theory}, because
such a waveguide supports only one fundamental mode (see Figure 2b).
The BS is realized using a directional coupler (DC), which is composed
of two waveguides. In the coupling region, the evanescent fields of
the two waveguide modes couple with each other and exchange energy.
As a result, two new coupling eigenmodes, the symmetric (Figure 2c)
and anti-symmetric (Figure 2d) superpositions of the two waveguide
modes, are generated. Owing to the different effective refractive
indices of these two modes, the beating of the two modes leads to
a BS-like function. By controlling the coupling strength, the amount
of output at the two waveguide ports (the splitting ratio) can be
tuned. Using the engineered waveguide gap, we obtained a coupling
profile with a splitting ratio of approximately 1:1.

\begin{figure}[htb]
\includegraphics[width=8cm]{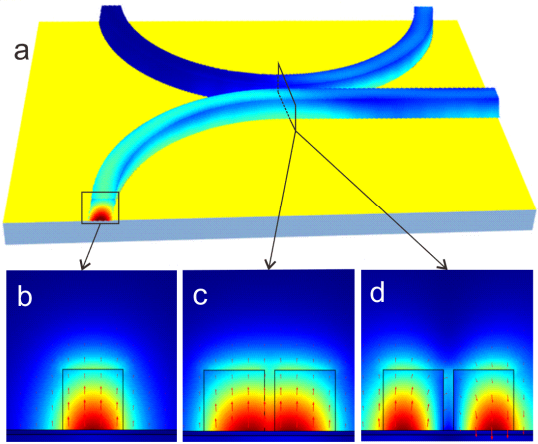} \caption{(a) Three-dimensional simulation of field distribution on our plasmonic
DC structure. (b) Field distribution in single mode plasmonic waveguide
with lateral size of 600nm$\times$600nm. (c) Field distribution of
symmetric eigenmode in coupling section. (d) Field distribution of
anti-symmetric eigenmode in coupling section. }
\end{figure}
The coupling efficiencies among our SPP circuit, the external source,
and the detectors were particularly crucial because the quantum signals
were weak (approximately 7,000 photons pairs per second in our experiment).
However, it is difficult to directly connect our plasmonic waveguide
to a single-mode optical fiber because its lateral-mode field area
is much smaller than that of the fiber (diameter $6.8\mu m$, 980HP,
Thorlabs Inc.). Therefore, we adopted the alternative adiabatic method
\cite{PBS,Zou} to excite the plasmons using fiber taper \cite{Dong,Tong}.
As Figure 1c shows, the photons in the fiber are adiabatically squeezed
into the microfiber via the taper region and coupled to the plasmon
waveguide when the microfiber approached the waveguide. Owing to the
high efficiency conversion and evanescent field coupling, the ideal
conversion efficiency might have been higher than 99\%. Under the
limitations imposed by the experimental conditions, the efficiency
of our fiber taper coupling system was estimated to be approximately
30\%. Importantly, the alignment direction of the fiber taper was
vertical to the collection fiber, thereby avoiding the collection
of directly scattered photons from the end of the fiber taper.

\begin{figure*}[htb]
\includegraphics[width=15cm]{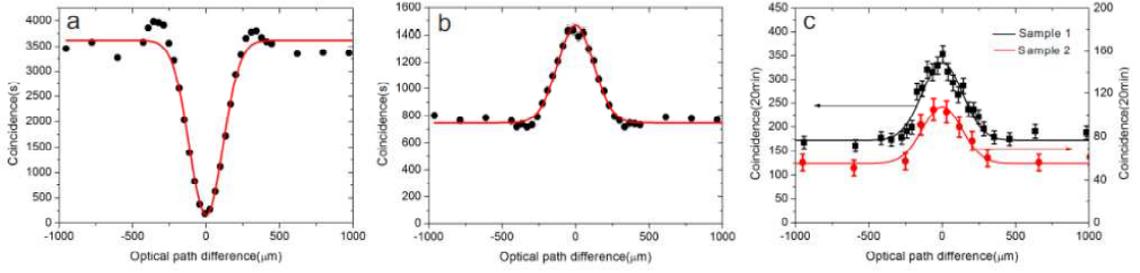} \caption{(a) HOM interference of the down-converted photon pairs measured using
a fiber 50/50 BS; the visibility was $95.5\pm1.0\%$ and the optical
coherence length was $162.6\pm5.0\mu m$. (b) Modified HOM interference
of the down-converted photon pairs measured using two fiber 50/50
BSs; the visibility was $96.5\pm3.1\%$, and the optical coherence
length was $173.9\pm5.7\mu m$. (c) Quantum interference of single
plasmons on DLSPPWs: For Sample 1, the visibility was $95.7\pm8.9\%$,
and the optical coherence length was $191.6\pm17.6\mu m$; for Sample
2, the visibility was $93.6\pm6.7\%$, and the optical coherence length
was $193.0\pm13.0\mu m$. All results are at the level of single photons.}
\end{figure*}

\textbf{Quantum-interference results.} The 1,550-nm quantum photon
pairs were generated via the spontaneous parametric down-conversion
(SPDC) \cite{Burn} process of a BBO crystal (Type-II phase matching,
non-collinear) pumped by a 775-nm-wavelength laser (Coherent Inc.;
see Figure 1a). The down-converted twin photons consisted of one photon
in the horizontal (H) polarization and one in the vertical (V) polarization.
The photons were separated into two paths, each of which contained
a prime reflector (PR), a half-wave plate (HWP, 1,550 nm), a long-pass
filter (LP; 830 nm), and a narrow-band filter (IF, 1,550 nm, 8.8 nm
FWHM). After these components, the two photons, which now had the
same polarization, were guided into two separate single-mode fibers.
One of the fiber couplers was installed on a motorized stage to adjust
the optical path.

As shown in Figure 3a, the indistinguishability of the produced photon
pairs was first characterized using a standard HOM interferometer
with a fiber BS. The dip represented the quantum interference of two
photons that arrived at the BS simultaneously, and the coherence of
the photons determined its width. The quantum-interference results
were fit using $N_{HOM}=C\cdot[1-V\cdot e^{-(\Delta\omega\cdot\Delta\tau)^{2}}]$\cite{Hong},
where $N_{HOM}$ is the measured coincidence count, $C$ is a fitting
constant, $V$ is the quantum-interference visibility, $\Delta\omega$
is the bandwidth of the photons, and $\Delta\tau$ is the optical
time delay. For perfect quantum interference of indistinguishable
photon pairs, the visibility should be unity. Here, we obtained a
visibility of $V=95.5\pm1.0\%$ and an optical coherence length of
$c/\Delta\omega=162.6\pm5.0\mu m$, where $c$ is the speed of light.
The deviation of the visibility from $100\%$ was attributed to the
polarization distortion of the photons during propagation in the fiber,
photon source variability , or both. We also tested the modified quantum
interferometer, in which photons from one output port were divided
using a second fiber BS and detected with two single-photon detectors.
Fitting the experimental results (Figure 3b) using the function $N_{M}=C\cdot[1+V\cdot e^{-(\Delta\omega\cdot\Delta\tau)^{2}}]$,
we obtained a visibility of $96.5\pm3.1\%$ and a coherence length
of $173.9\pm5.7\mu m$. These values were consistent with standard
HOM interference.

Finally, we observed the quantum interference of single plasmons using
the modified quantum interferometer, in which two single photons from
the fiber excited plasmon pairs in separate waveguides, and quantum
interference occurred in the coupling section. We sought to collect
the two plasmons from the two output ports and record their coincidence
with a standard HOM interferometer; to do so, we required two additional
fiber tapers to collect the signal. To avoid this requirement, we
simplified the experimental design by collecting the photons scattered
from P2 using an end-fire-coupled single-mode fiber. Using the second
fiber BS, we divided the collected photons into two ports and measured
the coincidence. Three samples were measured, and the visibilities
were $95.7\pm8.9\%$, $93.6\pm6.7\%$, and $93.1\pm16.5\%$. These
values are well above the classical limitation of $50\%$. The coherence
lengths of the plasmons were also calculated using the experimental
data, yielding $191.6\pm17.6\mu m$,$193.0\pm13.0\mu m$ and$146.4\pm9.4\mu m$,
which were similar values to those of the photons. Our results demonstrate
that although the electron is a fermion, a single plasmon (i.e., the
quasi-particle of a collective electron-density wave) acts as a boson.
The high visibility also suggests that plasmonic structures can be
used in QPICs.

\textbf{Discussion.} In this section, we address the second question:
what is the influence of loss on quantum interference visibility?
The inevitable loss of SPPs attenuates the amplitude of light; thus,
gain materials are often used to compensate for this loss. In addition,
the first-order coherence of the photons is destroyed during absorption
and re-emission processes. It is necessary to determine how this loss
influences second-order quantum interference visibility and under
what conditions these losses are tolerable. The following discussion
provides a detailed account of the two-photon quantum interference
of lossy channels based on our plasmonic DC structure.

The operation of a four-port DC can be described as follows:
\begin{equation}
\begin{pmatrix}b_{1}^{\dagger}\\
b_{2}^{\dagger}
\end{pmatrix}=\begin{pmatrix}r & t\\
t & r
\end{pmatrix}\begin{pmatrix}a_{1}^{\dagger}\\
a_{2^{\dagger}}
\end{pmatrix},
\end{equation}
where $a_{1}^{\dagger}$ and $a_{2}^{\dagger}$ as well as $b_{1}^{\dagger}$
and $b_{2}^{\dagger}$ are the creation operators of the input and
output boson particles, and $r$ and $t$ are the amplitudes of the
reflection and transmission coefficients. The output state of the
input twin-particle state $|1,1\rangle$ is
\begin{equation}
|\Phi\rangle_{out}=\sqrt{2}rt|2,0\rangle+\sqrt{2}rt|0,2\rangle+(r^{2}+t^{2})|1,1\rangle
\end{equation}
multiplied by a normalization factor. Here, we discard the terms that
represent the loss of one or two particles because only the coincidence
counts were recorded in the experiment. The probability of finding
two particles in the same mode (proportional to the HOM interference
visibility) is
\begin{equation}
P=\dfrac{4|rt|^{2}}{4|rt|^{2}+|r^{2}+t^{2}|^{2}}.
\end{equation}
For a lossless system, the DC is characterized by its classical transmission
and reflection coefficients, $|r|^{2}$ and $|t|^{2}$. Thus, designing
a DC with $|r|^{2}=|t|^{2}$ should optimize quantum interference.
However, for a lossy system, the structures and microscopic transport
process of the DC will determine the second-order quantum coherence.
In our DLSPPW DC, when plasmons were propagated in the coupling region
of the two waveguides, they were in coherent superpositions of symmetric
and anti-symmetric modes (see Figures 2c and 2d). The precise microscopic
losses can be included using the coefficients \cite{Loss}
\begin{eqnarray}
r=\dfrac{e^{in_{2}k_{0}L}}{2}(e^{i\mathrm{Re}(\Delta n)k_{0}L}e^{-\mathrm{Im}(\Delta n)k_{0}L}+1)\\
t=\dfrac{e^{in_{2}k_{0}L}}{2}(e^{i\mathrm{Re}(\Delta n)k_{0}L}e^{-\mathrm{Im}(\Delta n)k_{0}L}-1)
\end{eqnarray}
Here, $\Delta n=n_{1}-n_{2}$, where $n_{1(2)}$ is the effective
refractive index of the symmetric mode (the anti-symmetric mode),
$k_{0}$ is the wave vector in free space, and $L$ is the coupling
length. The imaginary portion of $n_{1(2)}$ corresponds to the propagation
loss of the plasmons and leads to a non-unitary operation matrix for
the DC.

By substituting Eqs. (4) and (5 ) into Eq. (3), we obtain $P$, which
is related to the loss difference between the two intermediate eigenmodes
($\propto \mathrm{Im}(\Delta n)$) and the coupling length $L$ . When
$L$ is sufficiently large, the energy in the eigenmode with higher
loss approximates 0 and can therefore be neglected compared with the
lower-loss eigenmode. In this case, $P$ decreases to 0.5, which corresponds
to a classical random process.

Figure 4 illustrates the relationships among $P$ and $L$ for a lossless
DC (black dots), our DLSPPW DC (blue dots), and a metal-strip DC (red
dots) in which we selected the $L$ that corresponded to a 50/50 splitter
. For a lossless DC, $P=1$ for any selected $L$. In our sample,
$P$ slowly decreased as $L$ increased. This result is because the
difference between $n_{1}$ ($1.318-0.00426i$) and $n_{2}$ ($1.150-0.00437i$)
is small; therefore, we were able to achieve a high interference visibility
for a small $L$. For the metal-strip DC used in \cite{Rei}, $P$
fastly decreased as $L$ increased because the difference between
$n_{1}$ ($2.036-0.02i$) and $n_{2}$ ($1.841-0.01i$) was much larger,
especially in the imaginary portions.

The influence of loss on quantum coherence defined the limitations
of lossy QPIC devices. The high-order quantum interference of photons
should be considered when designing integrated photonic components
because the microscopic processes of photons in these devices might
deviate from the expected unitary evolution.

In summary, we experimentally demonstrated that single plasmons can
be used as qubits to perform on-chip quantum information processing.
The discussion presented here regarding loss also introduces a platform
for using plasmonic structures to investigate the on-chip quantum-decoherence
phenomenon. Additional investigations should consider using single
plasmons as qubits to carry quantum information and achieve on-chip
linear optical computations or quantum simulations.

\begin{figure}[htb]
\includegraphics[width=8cm]{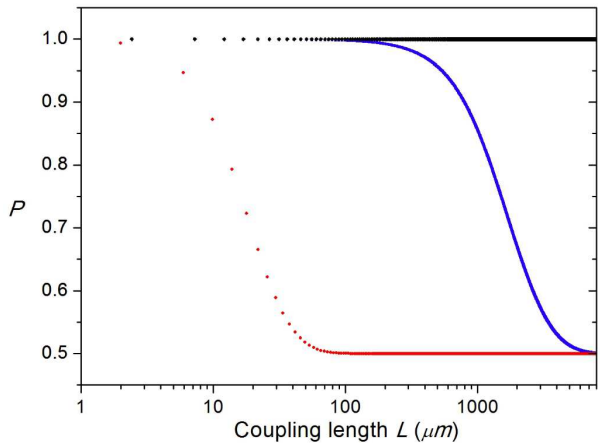} \caption{The relationship between $P$ and the coupling length $L$. The black,
blue, and red dots represent the theoretical calculations for a lossless
DC, our DLSPPW DC and a metal-strip DC \cite{Rei}, respectively.
$R$ decreases as $L$ increases and converges to 50\% for sufficiently
large $L$ in lossy DCs. Here, we used the $L$ values that corresponded
to a 50/50 splitter. }
\end{figure}

\section*{Acknowledgments}

This work was funded by NBRP (grant nos. 2011CBA00200 and 2011CB921200),
the Innovation Funds from the Chinese Academy of Sciences (grant no.
60921091), NNSF (grant nos.11374289, 10934006, 11374288, and 11104261),
and NCET. We thank Prof. Fang-Wen Sun, Bao-Sen Shi and Zheng-Wei Zhou
for useful discussion and Mrs Jun-Yi Xue from Qingdao No.2 High School
of Shandong Province for her help in optical measurement.


\begin{references}

\bibitem{Obrien08} Politi, A., Cryan, M. J., Rarity, J. G., Yu, S.
\& O'Brien, J. L. Silicaon-silicon waveguide quantum circuits, \emph{Science}
\textbf{320}, 646-649 (2008).

\bibitem{Obrien09} Politi, A., Jonathan, C. F. M., O'Brien, J. L.
Shor's Quantum Factoring Algorithm on a Photonic Chip. \emph{Science}
\textbf{325}, 1221 (2009).

\bibitem{Ebbesen03} Barnes, W., Dereux, A. \& Ebbesen, T. Surface
plasmon subwavelength optics. \emph{Nature} \textbf{424}, 824-830
(2003).

\bibitem{Sch} Schuller, J. \emph{et al}. Plasmonics for extreme light
concentration and manipulation. \emph{Nat. Mater}. \textbf{9}, 193-204
(2010).

\bibitem{PBS} Zou, C. L. \emph{et al}. Broadband integrated polarization
beam splitter with surface plasmon. \emph{Opt. Lett}. \textbf{36},
3630 (2011).

\bibitem{Polarizer} Dong, C. H., Zou, C. L., Ren, X. F., Guo, G.
C. \& Sun, F. W. In-line high efficient fiber polarizer based on surface
plasmon, \emph{Appl. Phys. Lett}. \textbf{100}, 041104 (2012).

\bibitem{Alt} Altewischer, E., van Exter, M. P. \& Woerdman, J. P.
Plasmon-assisted transmission of entangled photons. \emph{Nature}
\textbf{418}, 304-306 (2002).

\bibitem{energy} Fasel, S. \emph{et al}. Energy-time entanglement
preservation in plasmon-assisted light transmission. \emph{Phys. Rev.
Lett}. \textbf{94}, 110501 (2005).

\bibitem{ren06} Ren, X. F., Guo, G. P., Huang, Y. F., Li, C. F. \&
Guo, G. C. Plasmon-assisted transmission of high-dimensional orbital
angular-momentum entangled state. \emph{Europhys. Lett}. \textbf{76},
753-759 (2006).

\bibitem{Hong} Hong, C., Ou, Z. \& Mandel, L. Measurement of subpicosecond
time intervals between two photons by interference. \emph{Phys. Rev.
Lett}. \textbf{59}, 2044-2046 (1987).

\bibitem{Fuji} Fujii, G. \emph{et al}. Preservation of photon indistinguishability
after transmission through surface-plasmon polariton waveguide. \emph{Opt.
Lett}. \textbf{37}, 1535-1537 (2012).

\bibitem{Rei} Reinier W. H., Leo P. K. \& Valery Z. Quantum interference
in plasmonic circuits. \emph{Nat. Nanotechnol.} \textbf{8}, 719-722
(2013).

\bibitem{Bar} Barreiro, J. T., Wei, T. C. \& Kwait, P. G. beating
the channel capacity limit for linear photonic superdense coding.
\emph{Nat. Phys}. \textbf{4}, 282 (2008).

\bibitem{Liu} Liu, B. H. \emph{et al}. Experimental control of the
transition from Markovian to non-Markovian dynamics of open quantum
systems. \emph{Nat. Phys}. \textbf{7}, 931 (2011).

\bibitem{Ste} Stefano, L. Quantum simulation of decoherence in optical
waveguide lattices. \emph{Opt. Lett}. \textbf{38}, 4884-4887 (2013).

\bibitem{Gupta} Gupta, S. D. \&  Agarwal, G. S. Two-photon quantum interference in plasmonics:
theory and applications. \emph{Opt. Lett}. \textbf{39}, 390-393 (2014).

\bibitem{Klm} Knill, E., Laflamme, R. \& Milburn, G. J. A scheme
for efficient quantum computation with linear optics. \emph{Nature}
\textbf{409}, 46 (2001).

\bibitem{Lim} Lim, L. Y. \emph{et al}. Generalized Hong\textendash{}Ou\textendash{}Mandel
experiments with bosons and fermions.\emph{New J. Phys. }\textbf{7}\emph{,
155} \emph{(2005)}.

\bibitem{Kumar} Kumar, A. \emph{et al}. Dielectric-loaded plasmonic
waveguide components: Going practical. \emph{Laser Photonics Rev.}
1-14 (2013).

\bibitem{Kumar2} Kumar, A. \emph{et al}. Power monitoring in dielectric-loaded
surface plasmon-polariton waveguides. \emph{Opt. Exp}. \textbf{19},
2972-2978 (2011).

\bibitem{Kala} Kalavrouziotis,D., \emph{et al}. Active plasmonics
in true data traffic applications: Thermo-optic ON/OFF gating using
a silicon-plasmonic asymmetric MZI. \emph{IEEE Photon. Technol. Lett}.
\textbf{24}, 1036-1038 (2012).

\bibitem{Theory} Holmgaard, T., \& Bozhevolnyi, S. I. Theoretical
analysis of dielectric-loaded surface plasmon-polariton waveguides.
\emph{Phys. Rev. B} \textbf{75}, 245405 (2007).

\bibitem{Zou} Zou, C. L. \emph{et al}. Movable fiber-integrated hybrid
plasmonic waveguide on metal film. \emph{IEEE Photonic Tech. Lett.}
\textbf{24}, 434-436 (2012).

\bibitem{Dong} Dong, C. H. \emph{et al}. Coupling of light from an
optical fiber taper into silver nanowires. \emph{Appl. Phys. Lett}.
\textbf{95}, 221109 (2009).

\bibitem{Tong} Guo, X. \emph{et al}. Direct Coupling of Plasmonic
and Photonic Nanowires for Hybrid Nanophotonic Components and Circuits.
\emph{Nano Lett}. \textbf{9}, 4515 (2009).

\bibitem{Burn} Burnham, D. \& Weinberg, D. Observation of simultaneity
in parametric production of optical photon pairs. \emph{Phys. Rev.
Lett}. \textbf{25}, 84-87 (1970).

\bibitem{Loss} Peter, N. T. \emph{et al}. Integrated photonic sensing.
\emph{New J. Phys.} \textbf{13}, 055024 (2011).

\end{references}
\end{document}